\documentclass[a4paper,12pt]{amsart}

\usepackage{graphicx}
\usepackage{epsf}

\newtheorem{definition}{Definition}[section]

\newtheorem{theorem}{Theorem}[section]

\newtheorem{example}{Example}[section]

\begin{document}

\title[Density of first returns, periodic orbits and KS entropy]
{Density of first Poincar\'e returns, periodic orbits, and Kolmogorov-Sinai entropy\\ 
\today}
%

\author[Paulo R. F. Pinto, M. S. Baptista and Isabel S. Labouriau]
{Paulo R. F. Pinto$^{(1)}$, M. S. Baptista$^{(1)}$$^{(2)}$ and Isabel S. Labouriau$^{(1)}$}

\maketitle
\begin{center}
\footnotesize
$^{(1)}$CMUP - Centro de Matem\'atica da Universidade do Porto\\
Rua do Campo Alegre, 687, 4169-007 Porto, Portugal\\

\bigskip

$^{(2)}$Institute for Complex Systems and Mathematical Biology\\
King's College, University of Aberdeen\\
AB24 3UE Aberdeen, UK

\bigskip
\bigskip


\end{center}
\begin{abstract}
  It is known that unstable periodic orbits of a given map give
  information about the natural measure of a chaotic attractor.  In
  this work we show how these orbits can be used to calculate the
  density function of the first Poincar\'e returns.  The close
  relation between periodic orbits and the Poincar\'e returns allows
  for estimates of relevant quantities in dynamical systems, as the
  Kolmogorov-Sinai entropy, in terms of this density function.  Since
  return times can be trivially observed and measured, our approach to
  calculate this entropy is highly oriented to the treatment of
  experimental systems.  We also develop a method for the numerical
  computation of unstable periodic orbits.
\end{abstract}
\section{Introduction}
Knowing how often
a dynamical system returns to some place in phase
space is fundamental to understand dynamics. There is a well
established way
to quantify that: the {\bf first Poincar\'e return (FPR)}, which
measures how much time a trajectory of a dynamical system takes to
make two consecutive returns to a given region. Due to their
stochastic behaviour, given a return time it is not feasible to
predict the future return times
and for that reason one is usually interested in calculating the frequency
with which the Poincar\'e returns happen, the {\bf density of the first
  Poncar\'e returns} (DFP).

This work explains the existence of a strong relationship between
unstable periodic orbits (UPOs) and the first Poincar\'e returns in
chaotic attractors. Unstable orbits and first Poincar\'e returns have
been usually employed as a tool to analyse and characterise dynamical
systems. With our novel approach we can calculate how frequently returns
happen by knowing only a few unstable periodic orbits. Additionally,
such relation allows
us to easily estimate other fundamental
quantities of dynamical systems such as the Kolmogorov-Sinai entropy.

Our motivation to search for a theoretical and simple way of
calculating the distribution of Poincar\'e return times comes from the
fact that they can be simply and quickly accessible in experiments and
also due to the wide range of complex systems that can be
characterized by such a distribution. Among many examples,
in Ref. \cite{tito} the return times were used 
to characterize a experimental
chaotic laser, in Refs.  \cite{kantz,altmann} they were used to
characterize extreme events, in Refs. \cite{viana1,baptista} they were
used to characterize fluctuations in fusion plasmas, and in Ref.
\cite{marwan} a series of application to complex data analysis were
described.

In addition, relevant quantifiers of low-dimensional chaotic systems
may be obtained by the statistical properties of the FPR such as the
dimensions and Lyapunov exponents \cite{dimensao2,saussol} and the
extreme value laws \cite{todd2}.  For most of the rigorous results
concerning the FPR, in particular the form of the DFP \cite{hirata2},
one needs to consider very long returns to arbitrarily small regions
in phase space, a condition that imposes limitations into the real
application to data sets.

We first show how the DFP can be calculated from only a few UPOs
inside a finite region. Then, we explain how the DFP can be used to
calculate quantities as the Kolmogorov-Sinai entropy, even when only
short return times are measured in finite regions of the phase space.

Our work is organized as follows:  We first introduce the work of Ref.
\cite{GOY}, which relates the natural measure of a chaotic attractor
to the UPOs embedded in a chaotic attractor.  The measure of a chaotic
attractor refers to the frequency of visits that a trajectory makes to
a portion of the phase space.  This measure is called natural when it
is invariant for typical initial conditions.  This appears in
Sec.~\ref{defres}, along with the relevant definitions. In
Sec.~\ref{secDensity} we define $\rho(\tau,S)$ the density of first
Poincar\'e returns for a time $\tau$ to a subset $S$ of phase space
and we study the relation between the UPOs and this function.  This
can be better understood if we classify the UPOs inside $S$ as
recurrent and non-recurrent.  Recurrent are those UPOs that return
more than once to the subset $S$ before completing its cycle.
Non-recurrent are UPOs that visits the subset $S$ only once in a
period.  While in the calculation of the natural measure of $S$ one
should consider the two types of UPOs with a given large period inside
it, for the calculation of the DFP for a time $\tau$ one should
consider only non-recurrent UPOs with a period $\tau$.
Sec.~\ref{secCalc} is mostly dedicated to show how to calculate
$\rho(\tau,S)$ even when not all non-recurrent UPOs of a large period
are known. Such a situation typically arises when the time $\tau$ is
large. We have numerically shown that the error of our estimation
becomes smaller, the longer the period of the UPOs and the larger the
number of UPOs considered.

Throughout the paper we illustrate results by presenting the
calculations for the \textit{tent map}.  Finally, in
Sec.~\ref{secNumerical} we show numerical results on the
\textit{logistic map} that support our approach.  In particular, we
obtain numerical estimates of the Kolmogorov-Sinai entropy, the most
successful invariant in dynamics, so far.  The estimates are obtained
considering the density of only short first return times, as discussed
in Sec.~\ref{secKS}.  The UPOs of period $p$ are computed numerically
as stable periodic orbits of a system of $p$ coupled cells, a method
described in \ref{Hamilton}.

\section{Definitions and results}
\label{defres}
Consider a d-dimensional $C^2$ map of the form $x_{n+1}=F(x_n)$, where 
$x\in \Omega\subset R^{d}$ and $\Omega$ represents the phase space of the system. 
Consider $A\subset\Omega$ to represent a chaotic attractor. 
By chaotic attractor we mean an attractor that has at least one positive Lyapunov exponent. 

For a subset $S$ of the phase space and an initial condition $x_0$ in
the basin of attraction of $A$, we define $\mu(x_0,S)$ as the fraction
of time the trajectory originating at $x_0$ spends in $S$ in the limit
that the length of the trajectory goes to infinity. So,
\begin{equation}
\label{nmea}
\mu(x_0,S)=\lim_{n\rightarrow \infty}\frac{\sharp\{F^{i}(x_0)\in S,\ 0\leq i\leq n\}}{n}.
\end{equation} 

\begin{definition}
  If $\mu(x_0,S)$ has the same value for almost every $x_0$ (with
  respect to the Lebesgue measure) in the basin of attraction of $A$,
  then we call the value $\mu(S)$ the \textbf{natural measure of $S$}.
\end{definition}

\bigskip

For now we assume that our chaotic attractor $A$ has always a natural
measure associated to it, normalized to have $\mu(A)=1$.  In
particular this means that the attractor is ergodic\cite{GOY}.

We also assume that the chaotic attractor $A$ is mixing: given two
subsets, $B_1$ and $B_2$, in $A$, we have:
$$
\lim_{n\rightarrow \infty} \mu(B_1\cap F^{-n}(B_2))=\mu(B_1)\mu(B_2).
$$

In addition, we consider $A$ to be a hyperbolic set.

The eigenvalues of the Jacobian matrix of the $n$-th iterate, $F^{n}$,
at the $j$th fixed point $x_j$ of $F^{n}$ are denoted by
$\lambda_{1j},\lambda_{2j},...,\lambda_{uj},\lambda_{(u+1)j},...,\lambda_{dj}$,
where we order the eigenvalues from the biggest, in magnitude, to the
lowest and the number of the unstable eigenvalues is $u$. Let $L_j(n)$
be the product of absolute values of the unstable eigenvalues at
$x_j$.

Then it was proved by Bowen in 1972 \cite{bowen} and also by Grebogi,
Ott and Yorke in 1988 \cite{GOY} the following:

\begin{theorem}
\label{thegoy}
For mixing hyperbolic chaotic attractors, the natural probability
measure of some closed subset $S$ of the d-dimensional phase space is
\begin{equation}
\label{equationgoy}
\mu(S)=\lim_{n\rightarrow\infty}\sum_{x_j}L^{-1}_j(n),
\end{equation}
where the summation is taken over all the fixed points $x_j\in S$ of $F^n$.
\end{theorem}   
This formula is the representation of the natural measure in terms of
the periodic orbits embedded in the chaotic attractor. To illustrate
how it works let us take a simple example like the tent map:

\begin{example}
\label{tenda}
Let us consider $F:[0,1]\rightarrow [0,1]$ such that
$$
F(x)=\left\{\begin{array}{l}
2x,\ if\ x\in [0,1/2]\\
2-2x,\ if\ x\in ]1/2,1]
\end{array}\right.
$$

For this map there is only one unstable direction.  Since the absolute
value of the derivative is constant in $[0,1]$ we have
$L_j(\tau)=L(\tau)=2^\tau$.

For the tent map, periodic points are uniformly distributed in
$[0,1]$.  Using this fact together with some of the ideas of G.H.
Gunaratne and I. Procaccia \cite{GP}, it is reasonable to write the
natural measure of a subset $S$ of $[0,1]$ as:
\begin{equation}
\label{zero}
\mu(S)=\lim_{\tau\rightarrow \infty}\frac{N(\tau,S)}{N(\tau)},
\end{equation}
where $N(\tau,S)$ is the number of fixed points of $F^{\tau}$ in $S$
and $N(\tau)$ is the number of fixed points of $F^{\tau}$ in all space
$[0,1]$.  For this particular case we have $N(\tau)=L(\tau)=L_j(\tau)$
and so
$$
\mu(S)= \lim_{\tau\rightarrow
  \infty}\frac{N(\tau,S)}{N(\tau)}=\lim_{\tau\rightarrow
  \infty}\frac{N(\tau,S)}{L(\tau)}=\lim_{\tau\rightarrow
  \infty}\sum_{j=1}^{N(\tau,S)} \frac{1}{L_j(\tau)}
$$
and we obtain the Grebogi, Ott and Yorke formula.

\end{example}

\section{Density of first returns and UPOs}
\label{secDensity}

In this section we relate the DFP, $\rho(\tau,S)$, and the UPOs of a
chaotic attractor. We show in Eq. (\ref{mainresult}) that
$\rho(\tau,S)$ can also be calculated in terms of the UPOs but one
should consider in Eq. (\ref{equationgoy}) only the non-recurrent
ones.

\subsection{First Poincar\'e returns}
Consider a map $F$ that generates a chaotic attractor
$A\subset\Omega$, where $\Omega$ is the phase space. The \textit{first
  Poincar\'e return} for a given subset $S\subset \Omega$ such that
$S\cap A\neq\emptyset$ is defined as follows.

\begin{definition}
\label{deffpr}
A natural number $\tau$, $\tau>0$, is the \textbf{first Poincar\'e
  return} to $S$ of a point $x_0\in S$ if $F^{\tau}(x_0)\in S$ and
there is no other $\tau^*<\tau$ such that $F^{\tau^*}(x_0)\in S$.
\end{definition}

A trajectory generates an infinite sequence,
$\tau_1,\tau_2,...,\tau_i$, of first returns where $\tau_1=\tau$ and
$\tau_i$ is the first Poincar\'e return of $F^{n_i}(x_0)$ with
$n_i=\sum_{n=1}^{i-1}\tau_n$.

The subset $S'$ of points in $S\subset \Omega$ that produce FPRs of
length $\tau$ to $S$ is given by
\begin{equation}\label{Slinha}
S'=S'(\tau,S)=\left(F^{-\tau}(S)\cap S\right)-\bigcup_{0<j<\tau}\left(F^{-j}(S)\cap S\right).
\end{equation}

\subsection{Density function}
In this work, we are concerned with systems for which the DFP
decreases exponentially as the length of the return time goes to
infinity. Such systems have mixing properties and as a consequence we
expect to find $\rho(\tau,S)\approx\mu(S)(1-\mu(S))^{\tau-1}$, where
$(1-\mu(S))^{\tau-1}$ represents the probability of a trajectory
remaining $\tau-1$ iterations out of the subset $S$.  We are
interested in systems for which the decay of $\rho(\tau)$ is
exponential, i.e., $\rho(\tau)\propto e^{-\alpha\tau}$.

The usual way of defining $\rho(\tau,S)$, for a given subset $S\subset
\Omega$, is by measuring the fraction of returns to $S$ that happen
with a given length $\tau$ with respect to all other possible first
returns [see Eq. (\ref{densidade})]. It is usually required for a
density that
$$
\int \rho(\tau,S) d\tau = 1.
$$

In this work, we also adopt a more appropriate definition for
$\rho(\tau,S)$ in terms of the natural measure. We define the function
$\rho(\tau,S)$ as the natural measure of the set of orbits that makes
a first return $\tau$ to $S$ divided by the natural measure in $S$.
More rigorously, we have:

\begin{definition}
\label{one}
The \textbf{density function} of the first Poincar\'e return $\tau$
for a particular subset $S\subset \Omega$ such that $\mu(S)\ne 0$ is
defined as
\begin{equation}
\label{abcd}
\rho(\tau,S)=\frac{\mu(S')}{\mu(S)},
\end{equation}
where $S'=S'(\tau,S)\subset S$ is the subset 
of points that produce FPRs of length $\tau$ defined in Eq. (\ref{Slinha}).
\end{definition}

Even for a simple dynamical system as the tent map, the analytical
calculation of $\rho(\tau,S)$ is not trivial.  However, an upper bound
for this function can be easily derived as in the following example:

\begin{example}
\label{tendadens}
Consider the tent map defined in example \ref{tenda}, for which the
natural measure coincides with the Lebesgue measure $\lambda$, and let
$S\subset [0,1]$ be a non-trivial closed interval.

To have a return to $S$ we only need to know the natural number $n^*$
such that $F^{n^*}(S)=[0,1]$.  Since $F$ is an expansion, this natural
number always exists.  To find it when $\lambda(S)=\epsilon>0$, we
first solve the equation $2^{x^*}={1}/{\epsilon}$ and get
$x^*={-\log(\epsilon)}/{\log(2)}$, so we take
$n^*=\left[{-\log(\epsilon)}/{\log(2)}\right]+1$, where $[x]$
represents the integer part of $x$.  Then $n^*$ is an upper bound for
$\tau_{min}$, the shortest first return to $S$.

Most intervals $S$ of small measure have large values of $\tau_{min}$
and $\tau_{min}\approx n^*$ is a good approximation.  A sharper upper
bound for $\tau_{min}$ in $S$ is the lowest period of an UPO that
visits it.

The set $D=F^{-n^*}(S)\cap S\neq \emptyset$ represents the fraction of
points in $S$ that return to $S$ (not necessarily first return) after
$n^*$ iterations.  Using Eq. (\ref{abcd}) and since $S'\subset D$ we
have
$$
\rho(n^*,S)\leq\frac{\lambda(D)}{\lambda(S)}\leq\frac{\epsilon\frac{1}{2^{n^*}}}{\epsilon}=2^{-n^*}.
$$

It is natural to expect that for $\tau$ of the order of $n^*$ and
close to $\tau_{min}$ we have $\rho(\tau,S)\leq 2^{-\tau}$.

We can write this equation as $\rho(\tau,S)\leq
e^{(-\tau\log(2))}=e^{(-\tau\lambda_1)}$, where $\lambda_1=\log(2)$ is
the Lyapunov exponent for the tent map. In fact, in 1991, G. M.
Zaslavsky and M. K. Tippett

\cite{zas2}\cite{zas} presented one formula for the exact value of

$\rho(\tau,S)$. That result can only be valid under the same
conditions that we have used previously, i.e. $\tau\approx\tau_{min}$
and for most sets of sufficiently small measure $\epsilon$, so that
$\tau_{min}\approx n^*$.

\end{example}

\subsection{Density function in terms of recurrent and non-recurrent UPOs}
\label{nm}  

Since our chaotic attractor $A$ is mixing, the natural measure
associated with $A$ satisfies, for any subset $S$ of nonzero measure:
$$
\mu(S)=\lim_{\tau\rightarrow \infty}\frac{\mu(S\cap F^{-\tau}(S))}{\mu(S)}.
$$

We can write the right hand side of the last equation, for any 
positive $\tau$, in two terms:
\begin{equation}
\label{qq2}
\frac{\mu(S\cap F^{-\tau}(S))}{\mu(S)}=\frac{\mu(S')}{\mu(S)}+\frac{\mu(S^*)}{\mu(S)}
\end{equation}
with $S'$ as defined in Eq. (\ref{Slinha}) and where $S^*=S^*(S,\tau)$
is the set of points in $S$ that are mapped to $S$ after $\tau$
iterations but for which $\tau$ is not the FPR to $S$, so $S'\cup
S^*=(S\cap F^{-\tau}(S))$ and $S'\cap S^*=\emptyset$.

An UPO of period $\tau$ is \textit{recurrent} with respect to a set
$S\subset\Omega$ if there is a point $x_0\in S$ in the UPO with
$F^n(x_0)\in S$ for $0<n<\tau$.  In other words, its FPR is less than
its period. Thus, the UPOs in the set $S^*$ are all recurrent.  We
refer to them as the recurrent UPOs \textit{inside} $S$.

Associated with the recurrent UPOs in $S$ we define
\begin{equation}
\label{f1}
\mu_{R}(\tau,S)=\sum_j \frac{1}{L_j^R(\tau)}
\end{equation}
and  associated with the non-recurrent UPOs in $S$ we define
\begin{equation}
\label{f2}
\mu_{NR}(\tau,S)=\sum_j \frac{1}{L_j^{NR}(\tau)}
\end{equation}
where $L_j^R(\tau)$ and $L_j^{NR}(\tau)$ refer, respectively, to the
product of the absolute values of the unstable eigenvalues of
recurrent and non-recurrent UPOs of period $\tau$ that visit $S$.

Notice that, if $\mu(S)\ne 0$,
$$
\lim_{\tau\rightarrow\infty}\frac{\mu(S^*)}{\mu(S)}=\lim_{\tau\rightarrow\infty}\mu_R(\tau,S)
$$
and
\begin{equation}
\label{fff}
\lim_{\tau\rightarrow\infty}\frac{\mu(S')}{\mu(S)}=\lim_{\tau\rightarrow\infty}\mu_{NR}(\tau,S)
\end{equation}
since $\mu(S^*)/\mu(S)$ measures the frequency with which chaotic
trajectories that are associated with the recurrent UPOs visit $S$ and
$\mu(S')/\mu(S)$ measures the frequency with which chaotic
trajectories that are associated with the non-recurrent UPOs visit
$S$.

Comparing Eqs. (\ref{abcd}), (\ref{qq2}) and (\ref{fff}) we obtain the following:

\bigskip

\textbf{Main Idea:} \textit{For a chaotic attractor $A$ generated by a
  mixing uniformly hyperbolic map $F$, for a small subset $S\subset
  A$, generated by a Markov partition and such that the measure in $S$
  is provided by the UPOs inside it, we have that}
\begin{equation}
\label{mainresult}
\rho(\tau,S)\approx\mu_{NR}(\tau,S),
\end{equation}
\textit{for a sufficiently large $\tau$.}
\textit{Moreover,}
$$
\mu(S)=\lim_{\tau\rightarrow\infty}[\rho(\tau,S)+\mu_{R}(\tau,S)].
$$
\bigskip

A Markov partition is a very special splitting of the phase space. For
the purpose of better justifying Eq. (\ref{mainresult}), if a region
$C(\tau)$ belongs to a Markov partition of order $\tau$, then there is
a sub-interval $\tilde{C}(\tau)$ of $C(\tau)$ that after $\tau$
iterations is mapped exactly over $C(\tau)$.  Moreover, points inside
$\tilde{C}(\tau)$ make first returns to $C(\tau)$ after $\tau$
iterations. Then, $\mu_R(\tau,C(\tau))$=0. As a consequence, for
sufficiently large $\tau$ we can write that $\mu[C(\tau)] \rightarrow
\rho[\tau,C(\tau)]$.

But approximation (\ref{mainresult}) remains valid for a small
nonzero $\tau$. The reason for that is the following: Notice that from
the way Kac's lemma is derived (see Sec.~\ref{Mdfpr}), Eq.
(\ref{equationgoy}) can be written as
$$
\mu(S)=\frac{\int_{\tau_{min}}^{\infty}\rho(\tau,S)d\tau}{<\tau>},
$$
where $<\tau>$ represents the average of the FPRs inside $S$, since
$\int_{\tau_{min}}^{\infty}\rho(\tau,S)d\tau=1$. This equation
illustrates that any possible existing error in the calculation of
$\mu(S)$ by Eq. (\ref{equationgoy}) is a summation over all errors
coming from $\rho(\tau,S)$ for all values of $\tau$ that we are
considering. As shown in Ref. \cite{GOY}, $\mu(S)$ can be calculated
by Eq. (\ref{equationgoy}) using UPOs with a small and finite period
$p$. This period is of the order of the time that the Perron-Frobenius
operator converges and thus linearization around UPOs can be used to
calculate the measure associated with them.  As a consequence, if
$\mu(S)$ can be well estimated for $p\approx 30$ then $\rho(\tau,S)$
can be well estimated for $\tau<<p$. As we will observe, considering
$\tau$ small, of the order of $5$, we get a very good estimation for
$\rho(\tau,S)$.

In addition, we observe in our numerical simulation that $S$ does not
need to be a cell in a Markov partition but just a small region
located in an arbitrary location in $\Omega$.

\bigskip
 
We say that an UPO has FPRs associated with it if the UPO is
non-recurrent.  See that for every UPO there is a neighborhood
containing no other UPO with the same period.  If the UPO is
non-recurrent then all points inside a smaller neighborhood will
produce FPRs associated with this UPO in the sense that their FPR
coincides with the UPO's.  Consider $\tau_{min}$ as the shortest first
return in $S$.

\bigskip

\small
\textbf{Case  $\tau<2\tau_{min}$}
\normalsize

\bigskip

\begin{figure}
	\centering
        \includegraphics[width=12cm,height=9cm,viewport=0 -20 470
        440]{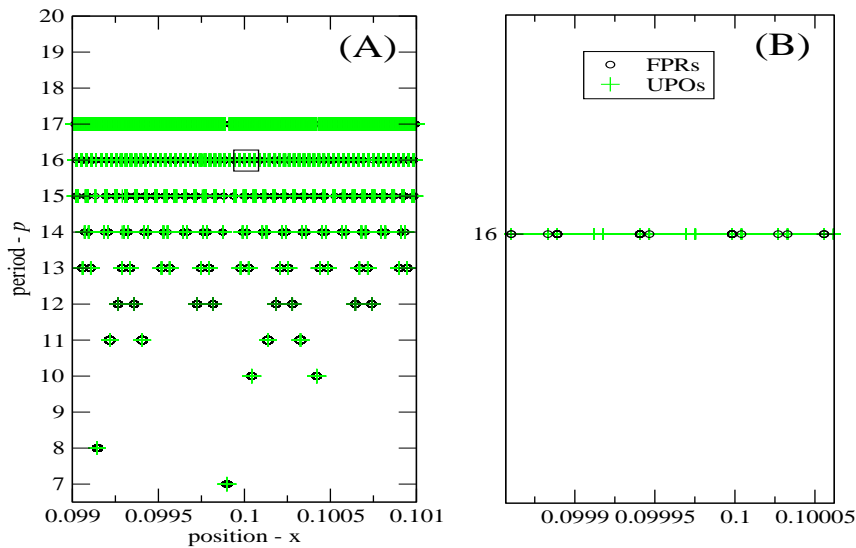}
        \caption{This picture shows some UPOs inside $S\subset[0,1]$
          and first Poincar\'e returns for the logistic map,
          [$x_{n+1}=4x_n(1-x_n)$]. In this example $\tau_{min}=7$. For
          $\tau<14$ all UPOs have FPRs associated with them. For
          $\tau\geq14$ (as in (B) for $\tau=16$) some UPOs are
          recurrent. Picture (B) is a zoom of picture (A).}
		\label{fprs}
\end{figure}

UPOs of period $\tau$ are non-recurrent.  This is illustrated in Fig.
\ref{fprs} (A), where $\tau_{min}=7$, for the logistic map ($c=4$).
In that picture we observe that for $\tau\leq14$ all FPRs are
associated with UPOs. Because of this fact $\mu(S^*)=0$ and then all
the chaotic trajectories that return to $S$ are associated with
non-recurrent UPOs.  So, $\rho(\tau,S)\approx\mu(S)$ and thus,
$\rho(\tau,S)\approx\mu_{NR}(\tau,S)$.

\bigskip

\small
\textbf{Case $\tau\geq 2\tau_{min}$}
\normalsize

\bigskip

We can have recurrent UPOs of period $\tau$, that do not have first
returns associated with them.  As a consequence $\mu(S^*)>0$ and
recurrent UPOs contribute to the measure of $S$.  This is illustrated
in Fig. \ref{fprs} (B), when $\tau=16$.

\section{How to calculate the density of first Poincar\'e returns}\label{secCalc}

A practical issue is how to calculate $\mu_{NR}(\tau,S)$. There are
two relevant cases: All UPOs can be calculated; only a few can be
calculated.

Assuming $\tau$ to be sufficiently small such that all UPOs of period
$\tau$ can be calculated and sufficiently large so that Eq.
(\ref{mainresult}) is reasonably valid, $\mu_{NR}(\tau,S)$ can be
exactly calculated and we can easily estimate $\rho(\tau,S)$ from
Eq.(\ref{mainresult}), using $\rho(\tau,S)\approx\mu_{NR}(\tau,S)$.

When $\tau$ is large then, typically, only a few UPOs can be
calculated. For this case, it is difficult to use Eq.
(\ref{mainresult}) to estimate $\rho(\tau,S)$ since there will be too
many UPOs. In order to calculate $\rho(\tau,S)$ using
$\mu_{NR}(\tau,S)$ we do the following. First notice that
\begin{equation}
\label{sss}
\mu(S)=\lim_{\tau\rightarrow \infty}(\mu_{NR}(\tau,S)+\mu_{R}(\tau,S)).
\end{equation}

Considering then $\tau$ sufficiently large we have that
$$
\mu(S)\approx\mu_{NR}(\tau,S)+\mu_{R}(\tau,S)
$$
which can be rewritten (using Eq. (\ref{mainresult}) which says that
$\rho(\tau,S)\approx\mu_{NR}(\tau,S)$, for finite $\tau$] as
\begin{equation}
\label{five}
\rho(\tau,S)\approx \mu(S)-\mu_{R}(\tau,S)=\mu(S)\left(1-\frac{\mu_R(\tau,S)}{\mu(S)}\right).
\end{equation}

This equation allows us to reproduce, approximately, the function
$\rho(\tau,S)$, for any sufficiently large $\tau$, only using the
estimated value of the quotient
$$
\frac{\mu_R(\tau,S)}{\mu(S)}
$$ 
that is easy to obtain numerically, since not all UPOs should be
calculated but just a few ones with period $\tau$. We discuss this in
\ref{calculo} below.

\subsection{How can we estimate $\mu_{R}(\tau,S)/\mu(S)$?}\label{calculo}

Considering a subset $S$ and fixing $\tau$, we calculate a number $t$
of different UPOs with period $\tau$ (say, $t=50$) inside $S$ (It is
explained in Sec.~\ref{Hamilton} how to calculate numerically UPOs
with any period of a given map). These UPOs are calculated from
randomly selected symbolic sequences for which the generated UPOs
visit $S$. See that, for example, in the tent map, for $\tau=10$ and
$S=[0,\frac{1}{8}]$, we may have $2^{10}/8$ UPOs inside $S$ and so,
here $50$ UPOs inside $S$ is, in fact, a very small number of UPOs.

Now, we separate all the $t$ UPOs that visit $S$ into recurrent and
non-recurrent ones and suppose that we have $r$ recurrent and $nr$
non-recurrent such that $r+nr=t$. So, $r$ and $nr$ depend on $t$ and
$S$. With these particular $r(t,S)$ recurrent UPOs we use Eq.
(\ref{f1}) and we obtain
$$
\tilde{\mu}_{R}[\tau,S,r(t,S)]=\sum_{j=1}^{r(t,S)}\frac{1}{L_j^R(\tau)}
$$
where $L_j^R(\tau)$ represents the product of the absolute values of
the unstable eigenvalues of the $j$th recurrent UPO within the set of
$r(t,S)$ recurrent UPOs. See that this quantity is not equal to
$\mu_R(\tau,S)$ since we are not considering all recurrent UPOs inside
$S$ but just a small number $r(t,S)$ of them. We do the same thing
with the $nr(t,S)$ non-recurrent UPOs and obtain the quantity
$\tilde{\mu}_{NR}[\tau,S,nr(t,S)]$.

Finally, we observe that, for a sufficiently large $t$, we have
$$
\frac{\tilde{\mu}_R[\tau,S,r(t,S)]}{\tilde{\mu}(\tau,S,t)}\approx\frac{\mu_R(\tau,S)}{\mu(S)},
$$
where
$\tilde{\mu}(\tau,S,t)=\tilde{\mu}_R[\tau,S,r(t,S)]+\tilde{\mu}_{NR}[\tau,S,nr(t,S)]$.
Therefore, with only a few UPOs inside $S$ we calculate an estimated
value for $\rho(\tau,S)$. This estimation is represented by $\rho_M$
and is given by 
\begin{equation}
{\rho}_M[\tau,S,r(t,S)]=\mu(S)\left( 1- \frac{\tilde{\mu}_R[\tau,S,r(t,S)]}{\tilde{\mu}(\tau,S,t)} \right)
\label{estima01}
\end{equation}

Notice that, for a large $\tau$ we will have more recurrent UPOs than
non-recurrent ones and therefore the larger $\tau$ is, the larger is
the contribution of the recurrent UPOs to the measure inside $S$.

\subsection{Error in the estimation}

To study how much our estimation in Eq. (\ref{estima01}) depends on
the number $t$ of UPOs, we first assume that if all UPOs are known,
the calculated distribution in Eq. (\ref{mainresult}) is ``exact'', or
in other words it has a neglectable error as when compared to the real
distribution provided by Eq.  (\ref{abcd}). 

Then, the error in Eq. (\ref{estima01}) will depend on the deviation
of the quotient
\begin{equation}
q_1=\frac{\tilde{\mu}_R[\tau,S,r(t,S)]}{\tilde{\mu}(\tau,S,t)}, 
\label{quotient1}
\end{equation}
calculated when only $t$ UPOs are known,  to the 
quotient 
 \begin{equation}
q_2=\frac{\tilde{\mu}_R[\tau,S,r(t=N(\tau,S),S)]}{\tilde{\mu}(\tau,S,t=N(\tau,S))}, 
\label{quotient2}
\end{equation}
\noindent
calculated when all the $N(\tau,S)$ UPOs are known.

Thus, the amount of error that our estimate [Eq. (\ref{estima01})] has as when 
compared to the ``exact'' value of $\rho$ (when all the
UPOs are known) can be calculated by
\begin{equation}
E[\tau,S,t]=\frac{|q_1-q_2|}{q_2}
\label{error}
\end{equation}
\noindent
which means that the quantity $E$ gives the amount of deviation, in a
scale from 0 to 1, of $\rho_M$ [Eq.  (\ref{estima01})] as when
compared to the ``exact'' value of $\rho$ [Eq. (\ref{mainresult})].
Notice that in Eq. (\ref{error}), the quantity 100$E$ corresponds to
the percentage of error that our estimation has.

\subsection{Uniformly distributed UPOs}

There is another way to estimate the value of $\rho(\tau,S)$ in terms
of the number of UPOs in a subset $S$ of a chaotic attractor $A$. We
define $N(\tau)$ as the number of fixed points of $F^{\tau}$ in $A$,
$N(\tau,S)$ as the number of fixed points of $F^{\tau}$ in $S$,
$N_{R}(\tau,S)$ as the number of fixed points of $F^{\tau}$ in $S$
whose orbit under $F$ is recurrent and $N_{NR}(\tau,S)$ as the number
of fixed points of $F^{\tau}$ in $S$ whose orbit under $F$ is
non-recurrent. Then, for a sufficiently large $\tau$ and for a
uniformly hyperbolic dynamical system for which periodic points are
uniformly distributed in $A$, we have
$$
\mu_{R}(\tau,S)\approx\frac{N_{R}(\tau,S)}{N(\tau)},\ \ \mu_{NR}(\tau,S)\approx\frac{N_{NR}(\tau,S)}{N(\tau)}.
$$
Using the previous approximations we can write
$$
\mu(S)\approx\frac{N_{R}(\tau,S)}{N(\tau)}+\frac{N_{NR}(\tau,S)}{N(\tau)}=\frac{N(\tau,S)}{N(\tau)}.
$$
By Eq. (\ref{mainresult}) we may write $\rho(\tau,S)\approx\mu_{NR}(\tau,S)$ and we have that
\begin{equation}
\label{six1}
\rho(\tau,S)\approx\mu(S)-\frac{N_{R}(\tau,S)}{N(\tau)}.
\end{equation}
which can be written as
\begin{equation}
\label{six}
\rho(\tau,S)\approx\mu(S)\left(1-\frac{N_{R}(\tau,S)}{N(\tau,S)}\right).
\end{equation}
Again, we have an expression with a quotient
$$
\frac{N_{R}(\tau,S)}{N(\tau,S)}
$$
that is, again, easy to obtain numerically by the same technique from
which $\mu_R/\mu$ can be estimated and therefore we can obtain an
estimation for $\rho(\tau,S)$, represented by $\rho_N$, by
\begin{equation}
{\rho}_N[\tau,S,r(t,S)]=\mu(S)\left( 1- \frac{r(t,S)}{t} \right)
\label{estima02}
\end{equation}
\noindent
where $r(t,S)$ represents the number of recurrent UPOs out of a total
of $t$ UPOs, exactly as previously defined.

\section{Kolmogorov-Sinai entropy}
\label{secKS}
In 1958 Kolmogorov introduced the concept of entropy into ergodic
theory and this has been the most successful invariant so
far\cite{pw}.  In this section we explain how to use the density of
first Poincar\'e returns to estimate the Kolmogorov-Sinai entropy
$H_{KS}$.

The exposition here does not aim to be rigorous, only to explain how
we have arrived at the numerical estimates for the logistic map of
Sec.~\ref{secNumerical}.  
which is a non-uniformly hyperbolic map.

It is known that\cite{periodo}
\begin{equation}
\label{123} 
N(\tau)\propto \exp( \tau H_{KS} ).
\end{equation} 

Consider $F$ as a dynamical system that has the following property:

$$
\frac{N_{NR}(\tau,S)}{N(\tau)}\approx\mu_{NR}(\tau,S)\approx \rho(\tau,S),
$$
for a sufficiently large $\tau$. For example, dynamical systems for
which periodic points are uniformly distributed on the chaotic
attractor $A$ have this property.

Considering the tent map and $S\subset [0,1]$ such that
$N_{NR}(\tau,S)=1$ (if there is more that one non-recurrent UPO
of period $\tau$
 inside $S$ we shrink $S$ to have only one), we have
$\rho(\tau,S)\approx\frac{1}{2^{\tau}}$ that agrees with example
\ref{tendadens}, for $\tau$ close to $\tau_{min}$ and for most
intervals $S$. For other non-uniformly hyperbolic systems as the
logistic  the H\'enon maps, this property holds in an
approximate sense and this approximation is better the larger $\tau$
is and the closer the interval $S$ is to a Markov partition.

Using the last approximation together with Eq. (\ref{123}) we may write
$$
\frac{N_{NR}(\tau,S)}{\rho(\tau,S)}\approx b\exp( \tau H_{KS}),
$$
for some positive constant $b\in R$. So, we have that
\begin{equation}
\label{xx}
H_{KS}\approx \frac{1}{\tau}\log\left(\frac{N_{NR}(\tau,S)}{b\rho(\tau,S)}\right)=\frac{1}{\tau}\log\left(\frac{N_{NR}(\tau,S)}{\rho(\tau,S)}\right)-\frac{\log(b)}{\tau}.
\end{equation}
We define the quantity $H(\tau,S)$ as
\begin{equation}
\label{zzz}
H(\tau,S)=\frac{1}{\tau}\log\left(\frac{N_{NR}(\tau,S)}{\rho(\tau,S)}\right)
\end{equation}
and then, for $b\geq 1$, it is clear that
$$
H_{KS}\approx\frac{1}{\tau}\log\left(\frac{N_{NR}(\tau,S)}{b\rho(\tau,S)}\right)\leq H(\tau,S),
$$
so $H(\tau,S)$ is a local upper bound for the approximation of $H_{KS}$, considering a sufficiently large $\tau$. 

Supposing that there is at least one non-recurrent UPO inside $S$, then for large  $\tau$ we have 
$\frac{N_{NR}(\tau,S)}{\rho(\tau,S)} >\!\!> b$, as $b$ is constant. 
Thus, the term
$$
\frac{1}{\tau}\log\left(\frac{N_{NR}(\tau,S)}{\rho(\tau,S)}\right)
$$ 
dominates the expression (\ref{xx}), for longer times.

\bigskip

This equation allows us to obtain an upper bound for $\rho(\tau,S)$.
See that $\rho(\tau,S)\leq N_{NR}(\tau,S)\exp(-\tau H_{KS})$ and if
$\tau\approx\tau_{min}$ then $N_{NR}(\tau,S)\approx 1$ and we obtain
$\rho(\tau,S)\leq \exp(-\tau H_{KS})$ as in example \ref{tendadens}.

\bigskip

Equation (\ref{zzz}) depends on the choice of the subset $S$ and is
then a local estimation for $H_{KS}$. To have a global estimate we
take a finite number, $n$, of subsets $S_i$ in the chaotic attractor
and make a space average as
\begin{equation}
\label{zzz2}
\frac{1}{\tau n}\sum_{i=1}^n\log\left(\frac{N_{NR}(\tau,S_i)}{\rho(\tau,S_i)}\right).
\end{equation}
Better results are obtained taking the average over pairwise disjoint
subsets $S_i$ that are well distributed over $A$.

When we consider $N_{NR}(\tau,S)=1$ this means that we have only one
non-recurrent UPO, with period $\tau$, inside $S$. In general, for
sufficiently small subsets, $S_i$, we may have $N_{NR}(\tau,S_i)=1\
\forall i$ and we obtain an approximation that only depends on the
density function of the first Poincar\'e returns
\begin{equation}
\label{zzz3}
H_{KS}\approx\frac{1}{\tau n}\sum_i\log\left(\frac{1}{\rho(\tau,S_i)}\right).
\end{equation}
An equation which can be trivially used from the experimental point of
view since we just need to estimate $\rho(\tau,S_i)$ and we do not
need to know the UPOs. For practical purposes, we consider in Eqs.
(\ref{zzz}), (\ref{zzz2}) and (\ref{zzz3}) that $\tau=\tau_{min}$.

\section{Numerical results}
\label{secNumerical}

\begin{figure}
	\centering
        \includegraphics[width=12cm,height=11cm,viewport=0 -20 440
        440]{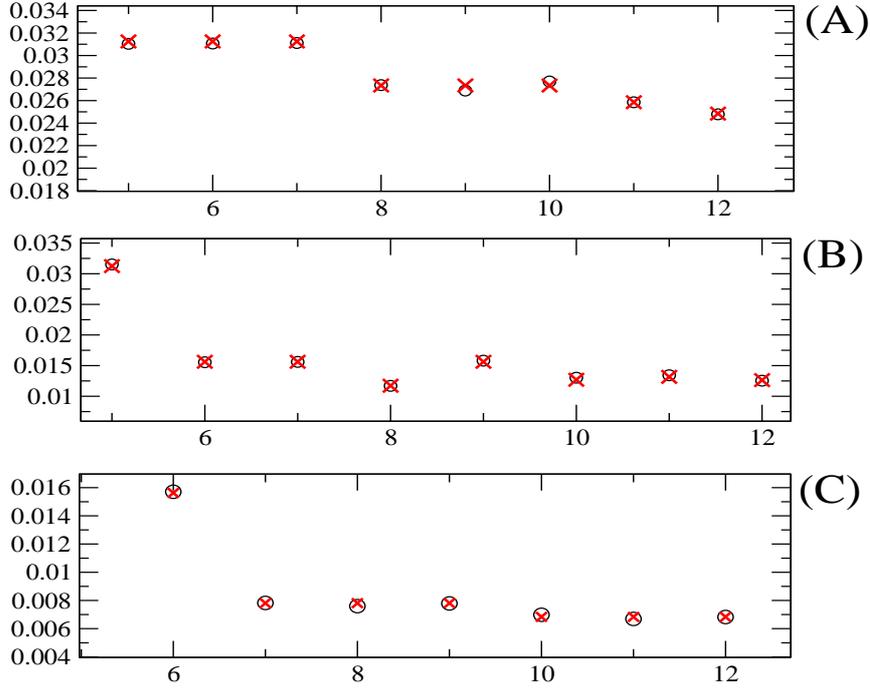}
        \caption{Density function of the FPRs, $\rho(\tau,S)$, as
          empty circles and the measure of the non-recurrent periodic
          orbits, $\mu_{NR}(\tau,S)$, as crosses, considering the
          following intervals: (A), $S=[0.3-0.05,0.3+0.05]$; (B),
          $S=[0.3-0.01,0.3+0.01]$; (C), $S=[0.3-0.005,0.3+0.005]$.}
		\label{ggg}
\end{figure}

\begin{figure}
  \centering
  \includegraphics[width=11cm,height=9cm,viewport=0 -20 420
  440]{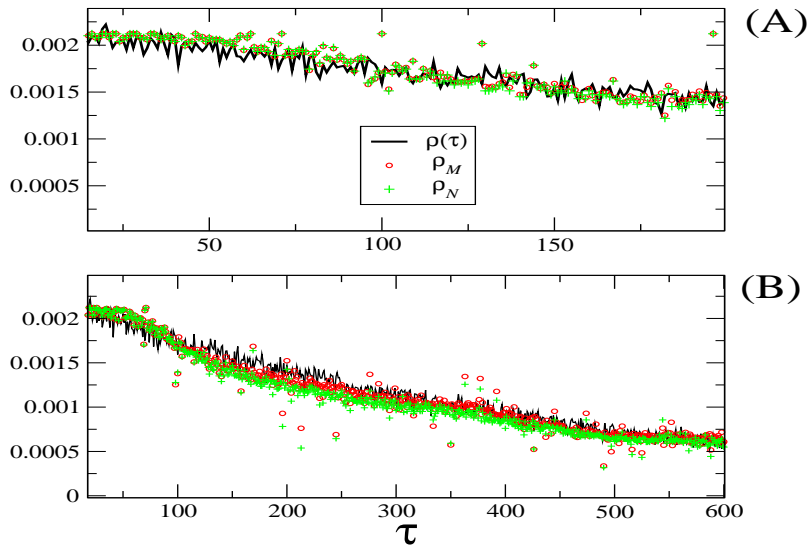}
  \caption{Red empty circles represent $\rho(\tau,S)$ estimated by Eq.
    (\ref{five}), green crosses estimated by Eq. (\ref{six}) and the
    black line calculated by Eq. (\ref{densidade}). Picture (B) is
    just a similar reproduction of (A) considering longer first return
    times. We consider 200 UPOs inside $S=[0.1-0.001,0.1+0.001]$, for
    each $\tau$.}
		\label{rhoapp}
\end{figure}

\begin{figure}
  \centering
  \includegraphics[width=12cm,height=10cm,viewport=0 -20 440
  450]{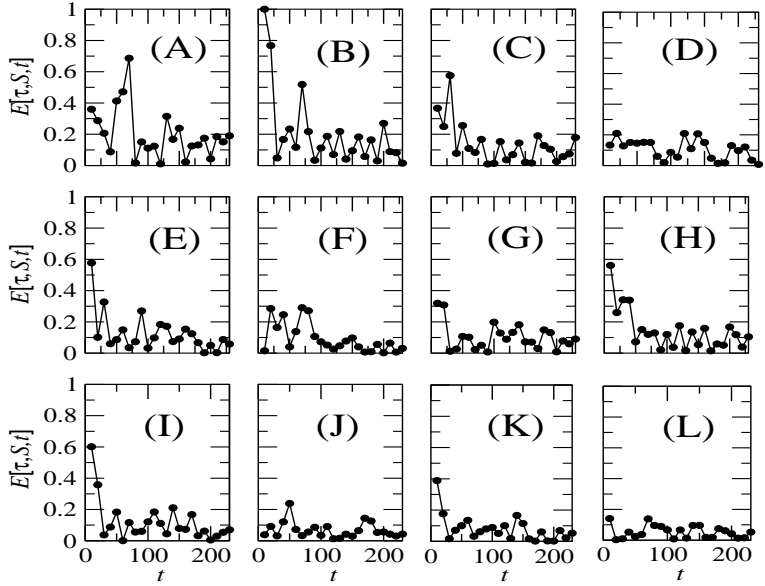}
  \caption{We show the quantity $E[\tau,S,t]$ with respect to the number
    $t$ of UPOs randomly chosen, for $\tau=9$ (A), $\tau=10$ (B),
    $\tau=11$ (C), $\tau=12$ (D), $\tau=13$ (E), $\tau=14$ (F),
    $\tau=15$ (G), $\tau=16$ (H), $\tau=17$ (I), $\tau=18$ (J),
    $\tau=19$ (K), and $\tau=20$ (L). The quantity $E$ gives the
    amount of deviation, in a scale from 0 to 1, of $\rho_M$ [Eq.
    (\ref{estima01})] as 
    compared to the ``exact'' value of
    $\rho$ [Eq. (\ref{mainresult})]. We consider an interval
    positioned at $x=0.04$ with size $\epsilon=0.02$.}
  \label{fig_erro}
\end{figure}

\begin{figure}
	\centering
        \includegraphics[width=12cm,height=10cm,viewport=0 -20 440
        450]{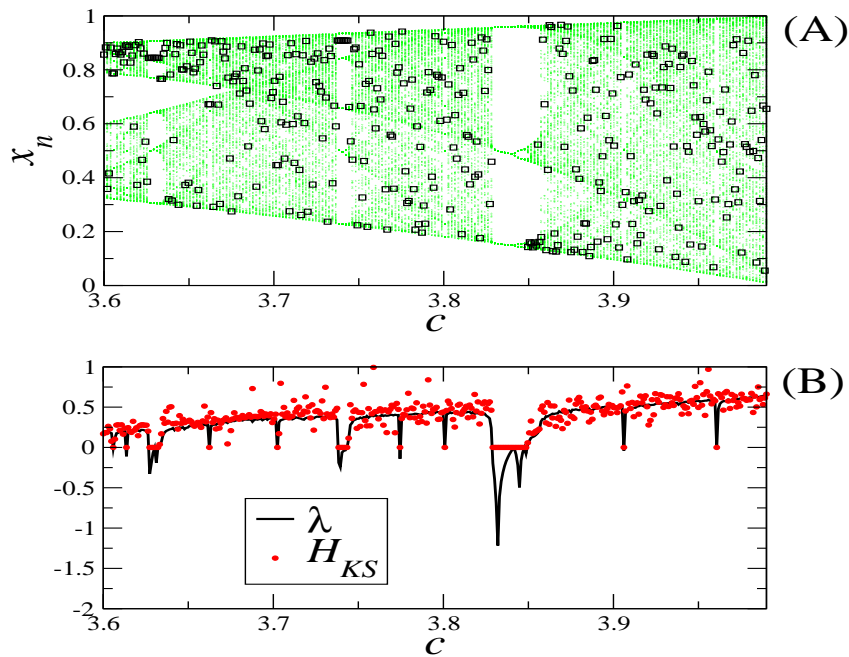}
        \caption{(A) A bifurcation diagram as points (light green) and the randomly
          chosen intervals as empty (black) squares. (B) Lyapunov exponent as
          line and filled circles representing the $H_{KS}$ entropy
          using Eq. (\ref{zzz}), for the logistic family. We consider
          400 values of $c$ and for each $c$ the size of the set $S$
          is $\epsilon=0.002$.}
        \label{hks}
\end{figure}

\begin{figure}
	\centering
        \includegraphics[width=11cm,height=10cm,viewport=0 -20 400
        450]{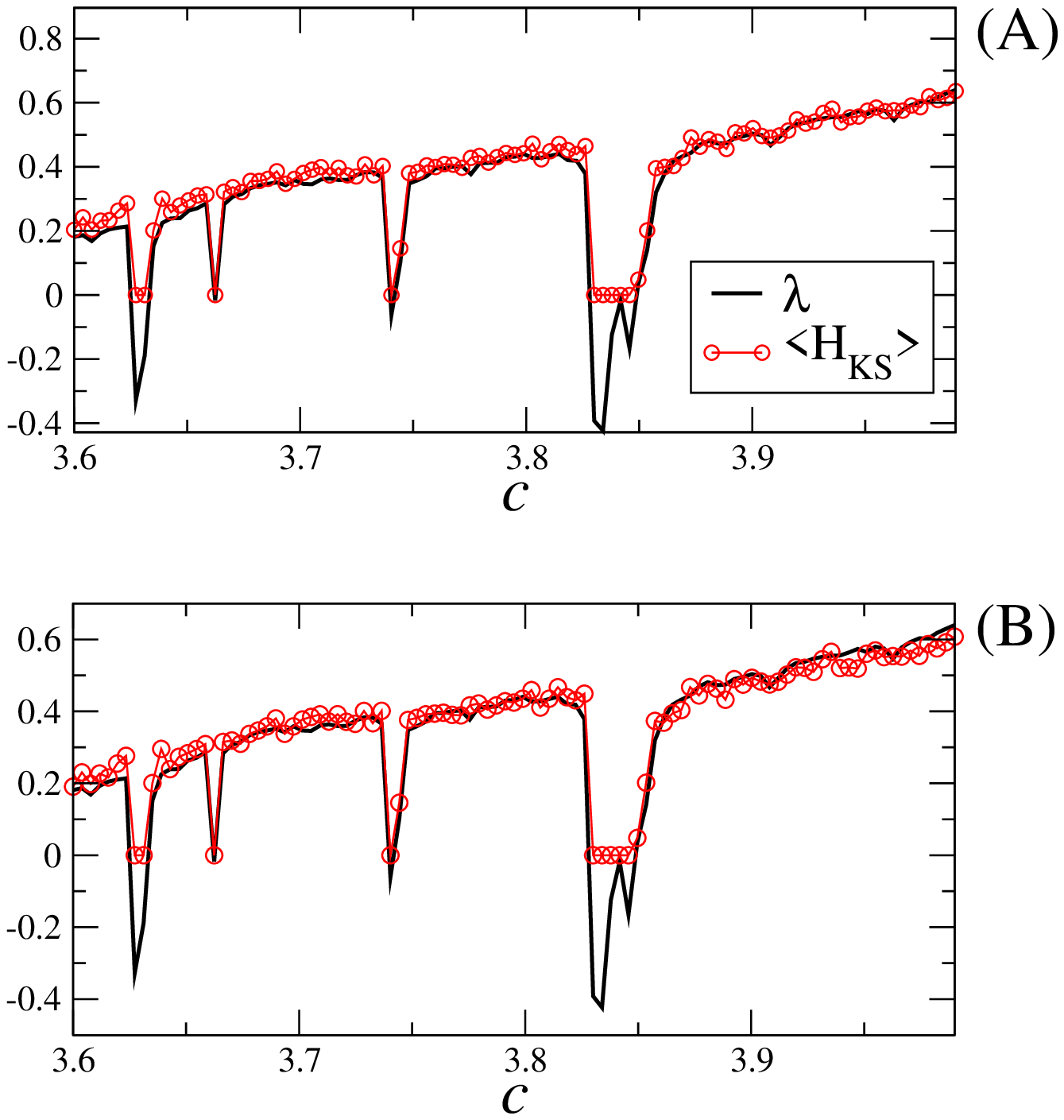}
        \caption{The Lyapunov exponent $\lambda$ as line and the
          aproximation of $H_{KS}$ entropy using Eqs. (\ref{zzz2}) and
          (\ref{zzz3}) as empty circles. (A), Eq. (\ref{zzz2}); (B),
          Eq. (\ref{zzz3}). In this simulation we consider 100 values
          of $c$ and for each $c$ we consider $40$ subsets $S_i$ each
          one with lenght $\epsilon=0.002$. A subset $S_i$ is picked
          only if $\tau_{min}\in[10,14]$.}
        \label{hks2}
\end{figure}


We illustrate our ideas with simulations on the logistic family $F:[0,1]\rightarrow[0,1]$ given by
\begin{equation}
\label{kkk}
F(x)=cx(1-x),
\end{equation}
were $c\in R$. There are many biological motivations to study this
family of maps\cite{Murray}. The maps that we obtain when the
parameter $c$ is varied have interesting mathematical properties. It
is therefore of relevant use for mathematical and biological study.
Moreover, for this family it is possible to compare the estimates made using all the UPOs to those using only some UPOs.

For most numerical simulations in this section we take $c=4$ in Eq.
(\ref{kkk}), for which the map is chaotic and the chaotic attractor is
compact.

\subsection{Calculating $\rho$ when all UPOs are known}

Figure \ref{ggg} shows the function $\rho(\tau,S)$ calculated by Eq.
(\ref{densidade}) and the values of $\mu_{NR}(\tau,S)$ calculated by
Eq. (\ref{f2}), for some subsets $S$. See that the DFP can be almost
exactly obtained if all the non-recurrent UPOs inside $S$ with period
$\tau$ can be calculated: In Sec.~\ref{secDensity} we concluded that
$\rho(\tau,S)\approx\mu_{NR}(\tau,S)$.

\subsection{Calculating $\rho$ when not all UPOs are known}

Figure \ref{rhoapp} shows the approximations for $\rho(\tau,S)$ using
Eqs. (\ref{estima01}) and (\ref{estima02}). In (B), comparing with
(A), we consider longer first return times. We only use Eqs.
(\ref{estima01}) and (\ref{estima02}) for $\tau>2\tau_{min}$.

\subsection{Error of our estimation when not all UPOs are known}

To numerically calculate the error [Eq. (\ref{error})] of our
estimation in Eq.  (\ref{estima01}), we only consider UPOs with a
period smaller than 20. The reason is because in order to calculate
the quotient $q_2$ in Eq. (\ref{quotient2}), all the UPOs must be
known. Considering larger periods than 20 would be computationally
demanding, even thought the proposed method to calculate UPOs is
capable of finding them all. 

It is also required that $\tau>2\tau_{min}$, once that to calculate
the quotient $q_1$ in Eq.  (\ref{quotient1}) there has to exist at
least one recurrent UPO within the set of $t$ UPOs considered, i.e.
$r \geq 1$.  Therefore, we need to choose the size of the interval
such that 20-2$\tau_{min}-1$ is sufficiently large, meaning an
interval for which $\tau_{min}$ is sufficiently smaller. We have
chosen $\epsilon$=0.02.

Since the error of our estimation is proportional to a quotient
between two quantities that depend on the number $r$ of recurrent
UPOs, it is advisable that one consider intervals for which a
reasonable number of recurrent UPOs are found, even when their period
is short (smaller or equal than 20). Such interval is positioned in
places were the natural measure is large. In the case of the logistic
map, these intervals are positioned either close to $x$=0 or $x=1$.
Therefore, we consider an interval positioned at $x=0.04$. From the
previous considerations, we consider that the interval has a size of
$\epsilon=0.02$.

In Fig. \ref{fig_erro}(A-I), we show the quantity $E[\tau,S,t]$ with
respect to the number $t$ of UPOs randomly chosen, for $\tau=9$ (A),
$\tau=10$ (B), $\tau=11$ (C), $\tau=12$ (D), $\tau=13$ (E), $\tau=14$
(F), $\tau=15$ (G), $\tau=16$ (H), $\tau=17$ (I), $\tau=18$ (J),
$\tau=19$ (K), and $\tau=20$ (L). 

The most important information from these figures is that as UPOs of
longer periods are considered [going from Fig. (A) to (L)], the error
$E$ of our estimation decreases in an average sense considering all
the values of $t$. Another relevant point is that the larger the
number $t$ of UPOs considered, the smaller the error. Notice that the
total number of UPOs of period $\tau$ is given by 2$^{\tau}$.
Therefore, looking at Fig.  \ref{fig_erro}(L), one can see that even
considering only  about 0.0009$\%$ of all the UPOs (10 UPOs, out of
a total of 2$^{20}$=1048576), the error of our estimation is smaller
than $14\%$ when compared to the ``exact'' value of $\rho$.

\subsection{Estimating the KS entropy}

In order to know how good our estimation for $H_{KS}$ is we use
Pesin's equality which states that $H_{KS}$ equals the sum of the
positive Lyapunov exponents, here denoted by $\lambda$.  For the
logistic map there is at most one positive Lyapunov exponent.

Figure \ref{hks} shows the approximation for the quantity $H_{KS}$
using Eq. (\ref{zzz}). See that Eq. (\ref{zzz}) only needs one subset
$S$ on the chaotic attractor to produce reasonable results. In this
numerical simulation we vary the parameter $c$ of the logistic family
and for each $c$ we use just one subset $S(c)$ randomly chosen [shown
in Fig. \ref{hks} (A)] but satisfying $\tau_{min}\in[10,14]$ so that
$\tau$ considered in Eq. (\ref{zzz}) is sufficiently large.

Finally, Fig. \ref{hks2} shows the global estimation for $H_{KS}$,
using the Eqs. (\ref{zzz2}) and (\ref{zzz3}), considering $40$
intervals $S_i$ for each value of $c$.  Recall that if $\lambda<0$,
then $H_{KS}=0$.

\subsection{Numerical work to find UPOs}
\label{Hamilton}

The analytical calculation of periodic orbits of a map is a difficult
task.  Even for the \textit{logistic map} it is very difficult to
calculate periodic orbits with a period as low as as four or five.  In
our numerical work we need to find unstable periodic orbits and, in
some cases, we need to find all different UPOs inside a subset of the
phase space, for a sufficiently large period. For that, we use the
method developed by Biham and Wenzel\cite{Ham}. They suggest a way to
obtain UPOs of a dynamical system with dimension $D$ using a
Hamiltonian, associated to the map, with dimension $ND$, where $N$ is
the number of UPOs with period $p$. The extremal configurations of
this Hamiltonian are the UPOs of the map. The force ${\partial
  H}/{\partial t}$ directs trajectories of the Hamiltonian to the
position of a UPO.

The Hamiltonian associated with the map gives a physical
interpretation of the problem but in some cases it is impossible to
know it. We propose a method with a similar interpretation that is
simpler in the sense that we do not need to know the Hamiltonian
associated with the map, just an array of $N$ coupled systems where
the linear coupling between nodes acts as the force directing the
network to possible periodic solutions of the dynamical system
concerned.

For this method we just need the force associated with the $i$th node,
described by $x^i$, and satisfying the Euler-Lagrange (E-L) equations:
$$
\frac{\partial}{\partial t}\frac{\partial L}{\partial \dot
  x^i}=\frac{\partial L}{\partial x^i},
$$
where $L$ is the Lagrangian associated with the map. We are interested
only in static extremum configurations of the Hamiltonian and
therefore the kinetic term will be neglected\cite{Ham}. This implies
$$
\frac{\partial L}{\partial x^i}=0
$$

We illustrate the numerical calculation of UPOs with arbitrary length
applying it to the logistic family. Because the static (E-L) equations
reproduce the map, we have
$$
\frac{\partial L}{\partial x_n^i}=x_n^{i+1}-cx_n^i(1-x_n^i).
$$
The force of the $i$ node will be given by
$$
F_i=-\frac{\partial L}{\partial x_n^i}=-x_n^{i+1}+cx_n^i(1-x_n^i).
$$

When the chain is in stable or unstable equilibrium (an extremum
static configuration of the Hamiltonian), $F_i=0$ for all $i$. To find
a specific extremum configuration of order $p$ of the Hamiltonian we
introduce an artificial dynamical system defined by
\begin{equation}
\label{systacop}
\frac{\partial x_n^i}{\partial t}=s_i F_i,\ i=1,...,p,
\end{equation}
where $s_i=\pm 1$ represents the direction of the force with respect
to the $i$th node. This equation is solved subject to the periodic
boundary condition $x^{p+1}=x^1$ and when the forces in all nodes
decrease to zero the resulting structure $x^i$ is simultaneously an
extremum static configuration and an exact $p$-periodic orbit of the
logistic map.  For $c=4$, if we take $s_i=-1\ \forall i$ then we
obtain the trivial periodic point $x_i=0\ \forall i$. The different
ways to write $s_i$ will give different UPOs. We may look at $s_i$ as
the representation of the orbit in a symbolic dynamics with
$\Sigma=\{-1,1\}$, taking the trivial partition on the logistic map,
i.e., $s_i=-1$ if $x_i\in [0,1/2]$ and $s_i=1$ if $x_i\in [1/2,1]$.

Equation (\ref{systacop}) is in fact an equation for a network of
coupled maps. The UPOs with period $p$ embedded in the chaotic
attractor can be calculated by finding the stable periodic orbits of
the following array of maps constructed with $i=1,...,p$ nodes
$x_n^i$, where every node is connected to its nearest neighbor as in
$$
x_{n+1}^i= x_n^i -cs_i[x_n^{i+1}-F(x_n^i)],
$$
with the periodic boundary condition $x_n^p=x_n^1$, where the term
$cs_i[x_n^{i+1}-F(x_n^i)]$ represents the Lagrangian force.

\section{Conclusions}\label{secConclusions}

In this work we propose two ways to compute the density function of
the first Poincar\'e returns (DFP), using unstable periodic orbits
(UPOs), where the first Poincar\'e return (FPR) is the sequence of
time intervals that a trajectory takes to make two consecutive returns
to a specific region.  In the first way, the DFP can be exactly
calculated considering all UPOs of a given low period. In the second
way, the DFP is estimated considering only a few UPOs. We have
numerically shown that the error of our estimation becomes smaller,
the longer the period of the UPOs and the larger the number of UPOs
considered.

The relation between DFP and UPOs allows us to compute easily an
important invariant quantity, the Kolmogorov-Sinai entropy.

For non-uniformly hyperbolic systems there exists some particular
subsets for which the UPOs that visit it are not sufficient to
calculate their measure of the chaotic attractor inside
it\cite{lai,bap}. For such cases our approach works in an approximate
sense, but it still provides good estimates as we have shown in our
simulations performed in the logistic map, a non-uniformly hyperbolic
system. In addition, the approaches shown in here were applied in ref.
\cite{chaos} to estimate the value of the Lyapunov exponent in the
experimental Chua's circuit and in the H\'enon map, both systems being
non-hyperbolic.

Our approach offers an easy way to estimate the KS entropy in
experiments, since one does not need to calculate UPOs, but rather
only to measure the DFP of trajectories that make shortest returns,
i.e. the quantity $\rho(\tau_{min},S)$.  These 
are the
most frequent trajectories, and as a consequence even if only a few
returns are measured, one can obtain a good estimation of
$\rho(\tau_{min},S)$.  More details of how to estimate the KS entropy
from experimental data can be found in Ref.  \cite{chaos}.

\section{Appendix}

\subsection{Measure and density in terms of FPRs}
\label{Mdfpr}
We calculate $\rho(\tau,S)$ also in terms of a finite set of FPRs by
\begin{equation}
\label{densidade}
\rho(\tau,S)=\frac{K(\tau,S)}{L(S)}
\end{equation}
where $K(\tau,S)$ is the number of FPRs with a particular length
$\tau$ that occurred in region $S$ and $L(S)$ is the total number of
FPRs measured in $S$ with any possible length.

\bigskip

We calculate $\mu(S)$ also in terms of FPRs by
\begin{equation}
\label{medida}
\mu(S)=\frac{L(S)}{n_L}
\end{equation}
where $n_{L}$ is the number of iterations considered to measure the
$L(S)$ FPRs and so $n_L=\sum_{n=1}^{L}\tau_n$ (see definition
\ref{deffpr}).

\bigskip

We define the average of the returns by
\begin{equation}
\label{x}
<\tau>=\frac{n_L}{L(S)}.
\end{equation}
Comparing Eqs. (\ref{medida}) and (\ref{x}), we have that
\begin{equation}
\mu(S)=\frac{1}{<\tau>}
\end{equation}
also known as Kac's lemma.

\bigskip

Acknowledgments: This work was supported by Funda\c c\~ao para a
Ci\^encia e a Tecnologia (FCT), by Centro de Matem\'atica da
Universidade do Porto (CMUP) and by the
Mathematics Department of Oporto University.


\begin{thebibliography}{99}



\bibitem{tito} F. T. Arecchi, A. Lapucci, R. Meucci, {\sl Experimental Characterization of Shilnikov Chaos by Statistics of Return Times}, Europhysics Letters, Vol. 6, Issue 8  (1988) 677--682.
%
\bibitem{kantz} M. S. Santhanam, H. Kantz, {\sl Return Interval Distribution of Extreme Events and Long-term Memory}, Physical Review E, Vol. 78, Issue 5  (2008) 051113.
%
\bibitem{altmann} E. G. Altmann; H. Kantz, {\sl Recurrence Time Analysis, Long-term Correlations, and Extreme Events}, Physical Review E, Vol. 71, Issue 5  (2005) 056106.
%
\bibitem{viana1} Z. O. Guimar\~aes, I. L. Caldas, R. L. Viana, {\sl Recurrence Quantification Analysis of Electrostatic Fluctuations in Fusion Plasmas}, Physics Letters A, Vol. 372, Issue 7  (2008) 1088--1095.
%
\bibitem{baptista} M. S. Baptista, I. L. Caldas, M. V. A. P. Heller, A. A. Ferreira, {\sl Recurrence in Plasma Edge Turbulence}, Phys. Plasmas, 8 4455 (2001).
%
\bibitem{marwan} N. Marwan, A. Facchini, M. Thiel, {\sl 20 Year of Recurrence Plots: Perspectives for a Multi-purpose Tool of 
Nonlinear Data Analysis}, European Physical Journal-Special Topics, Vol. 164 (2008) 1--2.

\bibitem{dimensao2} J. B. Gao, {\sl Recurrence Time Statistics for Chaotic Systems and Their Applications}, Phys. Rev. Lett. 83 (1999) 3178--3181.
%
%
\bibitem{saussol} B. Saussol, S. Troubetzkoy, S. Vaienti, {\sl Recurrence, dimensions and Lyapunov exponents}, J. of Stat. Phys. 106 (2002) 623--634.
%
%
%
%
%
\bibitem{todd2} A. C. M. Freitas, J. M. Freitas, M. Todd, {\sl Hitting Time Statistics and Extreme Value Theory}, arXiv:0804.2887.
%
%
%
\bibitem{hirata2} M. Hirata, B. Saussol, S. Vaienti, {\sl Statistics of return times: a general framework and new applications}, Comm. Math. Phys. 206 (1999) 33--55.
%
%

%
\bibitem{GOY} C. Grebogi, E. Ott, J. A. Yorke, {\sl Unstable periodic orbits and the dimensions of multifractal chaotic attractors}, Physical Review A 37 (1988) 1711--1724.
%
\bibitem{bowen} R Bowen, {\sl Periodic Orbits for Hyperbolic Flows}, American Journal of Mathematics 94 (1972) 1--30.
%



%
\bibitem{GP} G. H. Gunaratne, I. Procaccia, {\sl Organization of Chaos}, Phys. Rev. Lett.  59 (1987) 1377--1380.
%
\bibitem{zas2} G. M. Zaslavsky, M. K. Tippett, {\sl Connection between Recurrence-Time Statistics and Anomalous Transport}, Phys. Rev. Lett.  67 (1991) 3251--3254.
%
\bibitem{zas} G. M. Zaslavsky, {\sl Chaos, fractional kinetics, and anomalous transport}, Physics Reports 371 (2002) 461-580.
%

%
\bibitem{pw} Peter Walters, {\sl An Introduction to Ergodic Theory}, Springer, GTM number 79 (1981).
%
\bibitem{periodo} Ya. G. Sinai, {\sl Classical dynamic systems with countably-multiple Lebesgue spectrum}, Izv. Akad. Nauk SSSR, Ser. Mat. 30 1966 15--68. 
%
%
%
\bibitem{Murray} J. D. Murray, {\sl Mathematical Biology}, Springer, Biomathematics Texts number 19 (1993).
%
%
\bibitem{Ham} O. Biham, W. Wenzel, {\sl Characterization of Unstable Periodic Orbits in Chaotic attractors and Repellers}, Phys. Rev. Lett.  63 (1989) 819--822.

\bibitem{lai} Y.-C. Lai, Y. Nagai, C. Grebogi, {\sl Characterization of the Natural Measure by Unstable Periodic Orbits in Chaotic Attractors} Phys. Rev. Lett. 79 (1997) 649--652.
%
\bibitem{bap} M. S. Baptista, S. Kraut, C. Grebogi, {\sl Poincar\'e Recurrence and Measure of Hyperbolic and Nonhyperbolic Chaotic Attractors}, Phys. Rev. Lett. 95 094101 (2005).
%
\bibitem{chaos} M. S. Baptista, D. M. Maranh\~ao, J. C. Sartorelli, {\sl Dynamical estimates of chaotic systems from Poincar\'e recurrences}, Chaos, 19 043115 (2009).
%
\end{thebibliography}
\end{document}